\documentclass{iaus}
\usepackage{graphicx}
\usepackage{natbib}
\newcommand{\aj}{\textit{AJ}}
\newcommand{\aap}{\textit{A\&A}}
\newcommand{\apss}{\textit{Ap\&SS}}
\newcommand{\apj}{\textit{ApJ}}
\newcommand{\apjl}{\textit{ApJ}}
\newcommand{\apjs}{\textit{ApJS}}
\newcommand{\mnras}{\textit{MNRAS}}
\newcommand{\msait}{\textit{MemSAIt}}
\newcommand{\nat}{\textit{Nature}}
\newcommand{\pasp}{\textit{PASP}}

\title[IAU 266.~~Globular clusters as stellar evolution laboratories] 
{Globular clusters as laboratories \\ for stellar evolution}

\author[M. Catelan, A. A. R. Valcarce, \& A. V. Sweigart]   
{M\'arcio Catelan,$^1$ Aldo A. R. Valcarce,$^1$
\and Allen V. Sweigart$^2$}

\affiliation{
$^1$Pontificia Universidad Cat\'olica de Chile,  
Departamento de Astronom\'{i}a y Astrof\'{i}sica, \\ 
Av. Vicu\~na Mackenna 4860,
782-0436 Macul, Santiago, Chile \\ 
email: {\tt [mcatelan, avalcarc]@astro.puc.cl} \\[\affilskip]
$^2$NASA Goddard Space Flight Center, Exploration of the Universe Division, Code 667,\\ Greenbelt, MD 20771, USA \\
email: {\tt allen.v.sweigart@nasa.gov}\\
}

\pubyear{2009}
\volume{266}  
\pagerange{1-12}
\jname{Star Clusters: Basic Galactic Building Blocks}
\editors{R. de Grijs \& J. R. D. L\'epine, eds.}
\begin{document}

\maketitle

\begin{abstract}
Globular clusters have long been considered the closest approximation to a physicist's
laboratory in astrophysics, and as such a near-ideal laboratory for (low-mass) stellar 
evolution. However, recent observations have cast a shadow on this long-standing paradigm, 
suggesting the presence of multiple populations with widely different abundance patterns, 
and~-- crucially~-- with widely different helium abundances as well. In this review we 
discuss which features of the Hertzsprung-Russel diagram may be used as helium 
abundance indicators, and present an overview of available constraints on the 
helium abundance in globular clusters. 


\end{abstract}

\firstsection 
\section{Introduction}\label{sec:intro}
In the words of \citet{sm01}, ``globular clusters [GC's] are the closest approximation 
to a physicist's laboratory in astrophysics.'' Not only do these spheroidal stellar 
agglomerations contain up to several million stars, all at the same distance from us, 
but their stars have also been thought to have all been born at the same instant, 
from a cloud with homogeneous chemical composition. According to this canonical 
paradigm, GC's represent the best examples of a simple stellar population. As
a result, GC's have been extensively used to place constraints  
on key ingredients of canonical stellar evolution models, such as the mixing length
parameter of convective energy transport theory \citep[e.g.,][]{rpea02,ffea06}. 
In addition, GC studies have played an important role in the field of particle 
physics, with measured color-magnitude diagram (CMD) properties being used to 
place some of the stringest constraints available on several dark matter candidates
and other particle physics parameters that play a key role in the physics beyond 
the so-called ``standard model'' of particle physics~-- including, e.g., the magnetic 
moment of the neutrino, the mass of the axion, and the cross sections of weakly 
interacting massive particles \citep[e.g.,][and references therein]{gr96,gr00,gr08}.
 
However, recent observations suggest that GC's may not play as reliable a role 
as astrophysical laboratories as previously believed. For one, large variations
in some light element abundances, particularly O, Na, Al, and Mg (but, importantly, 
not in the iron-peak or other $\alpha$-capture elements), have been found in virtually
all GC's for which suitable spectroscopic data have been obtained \citep[e.g.,][and 
references therein]{ecea09}, unlike what is seen among metal-poor field stars 
\citep{rgea00}. For another, recent {\em Hubble Space Telescope} observations
\citep[e.g.,][and references therein]{gp09a,gp09b} have revealed the presence of multiple 
stellar populations in at least some of the more massive GC's in our galaxy, including
multimodal main sequences (MS's) and subgiant branches (SGB's), possibly connected with
multimodal horizontal branches (HB's)~-- the latter having been known since much earlier 
\citep[see][for a recent review]{mc08}. 

While at least some GC CMD's do still reveal exquisitely tight sequences down to the 
bottom of the MS, with no evidence of multiple populations 
\citep[or even for sizeable samples of binary stars;][]{ddea08}, 
the fact that several GC's do show split sequences
strongly suggests that GC's do not all represent the simple stellar populations that they
were once thought to represent. Instead, they appear to be chemically complex entities 
whose stars do not form all at the same time, undergoing instead several bursts of star
formation which progressively contaminate the medium before the formation of the next
stellar generation in the cluster. 

Of particular importance, in this regard, is the possibility that each stellar generation
may increase the helium content of the medium before the next generation forms. Helium is
the second most abundant element in the Universe, and as such, changes in 
$Y$ may dramatically change the stellar evolutionary paths and associated 
timescales. In this sense, while helium enrichment is a consistent prediction of different 
scenarios for the abundance variations that are seen in GC's, the amount of 
He enhancement differs markedly from one model to the next \citep[e.g.,][]{amea09}. 
Importantly, while the observed MS splits in some GC's indicate very large levels of 
He enrichment, with the He abundance in the most extreme populations reaching values as 
high as about 40\% by mass \citep{jn04,fdea05,gpea05,gpea07}, it is not 
straightforward to produce such large levels of He enrichment on the basis of  
available chemical enrichment scenarios. In this sense, and given that He enhancements 
among GC stars have now been suggested to be the rule, rather than the exception 
\citep[with some GC's~-- those with purely blue HB's~-- possibly lacking
first generation, non-He-enriched stars altogether; e.g.,][]{dac08}, it is important 
to check all possible signatures of He enhancement, based on the observed
properties of GC stars, in order to constrain the He enrichment scenario. This is a 
crucial task to help establish the extent to which at least some GC's may still be safely 
used as laboratories of low-mass stellar evolution. On the other hand, the heavily 
contaminated GC's will still continue to play an important role, though as 
laboratories of the evolution (and ejecta produced by) more massive stars. 

Since He abundance measurements are only possible for hot stars, which are generally 
lacking in GC's, most of the available $Y$ estimates use 
indirect techniques, based on their CMD properties. In this sense, in the next 
section we first review empirical determinations of the He abundance in GC stars, 
and then discuss the impact of He abundance variations among several GC observables,
based on a new set of evolutionary tracks for canonical and He-enhanced compositions 
\citep{av10}, along with the results that have been obtained using some of these 
observables. Finally, in \S\ref{sec:conclusions} we present our conclusions.

\firstsection 
\section{Helium abundance measurements for GC's}\label{ref:CMD} 
\subsection{Direct Methods}\label{sec:direct} 
Most GC's lack sufficiently hot stars for direct He abundance 
measurements.\footnote{To be sure, a He~{\sc i} line is found in near-IR spectra of cool 
red giants at 10,830~\AA, but this is due to chromospheric emission, and is thus 
a much better indicator of mass loss than it is of He abundance proper 
\citep[e.g.,][and references therein]{adea09}.} 
Even for
those which do contain sufficiently hot stars, primarily on the blue HB, the results 
of direct He abundance measurements can be strongly affected by diffusion effects, 
which can dramatically lower the photospheric He abundance for HB stars hotter than 
11,500~K \citep[e.g.,][]{bb03,smea03}.\footnote{This corresponds to 
the so-called ``Grundahl jump'' \citep{fgea99}.}  
Conversely, at the hot end of the HB, very 
high photospheric He abundances have indeed been measured, which several authors 
have interpreted in terms of the ``late flasher'' scenario for the origin of these
hot stars. In this scenario, stars that lose too much mass while on the red
giant branch (RGB) may ignite He not at the RGB tip, but rather as they evolve 
towards the white dwarf region of the CMD, undergoing extensive mixing in the process. 
Their photospheres may thus become dramatically enriched in He, to levels that can be 
much higher than even the highest levels that have been suggested under the ``primordial'' 
(multiple-populations) scenario  
\citep[e.g.,][and references therein]{smea04,smea07,scea09}. 

This leaves us with blue HB stars situated between the blue edge of the instability 
strip, at around 7200~K, and the Grundahl jump, at 11,500~K. Very recently, \citet{svea09}
carried out the first detailed spectroscopic He abundance measurement for 
HB stars in this temperature range. Based on UVES@VLT spectra of cool blue HB stars 
in NGC~6752, they found that ``... all our targets... have a homogeneous He 
content with a mean value $Y = 0.245 \pm 0.012$, compatible with the most recent 
measurements of the primordial He content of the Universe.'' Note that NGC~6752, as 
a cluster whose HB is comprised solely of blue HB stars, might have been viewed 
as an object lacking first-generation (i.e., non-He-enriched) stars~-- but these 
new measurements show that even clusters having entirely blue HB's likely have at 
least some stars with primordial He.

\begin{figure}[t]
\begin{center}
 \includegraphics[width=3.57in]{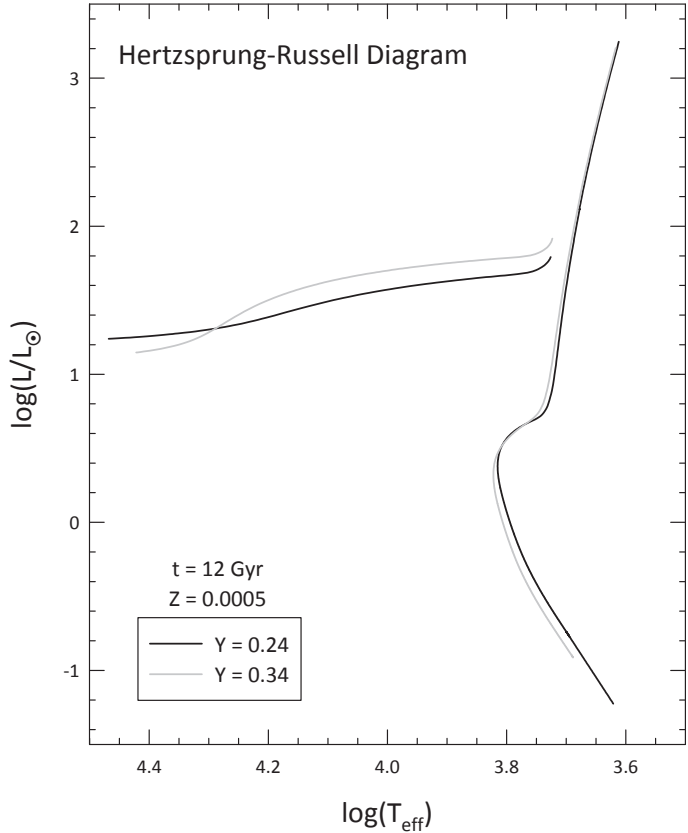} 
 \caption{Isochrones for an age of 12~Gyr, metallicity $Z = 0.0005$, and two different initial 
  helium abundances: $Y = 0.24$ ({\em black lines}) and $Y = 0.34$ ({\em gray lines}).}
   \label{fig:HRD}
\end{center}
\end{figure}

\begin{figure}[ht]
\begin{center}
 \includegraphics[width=3.57in]{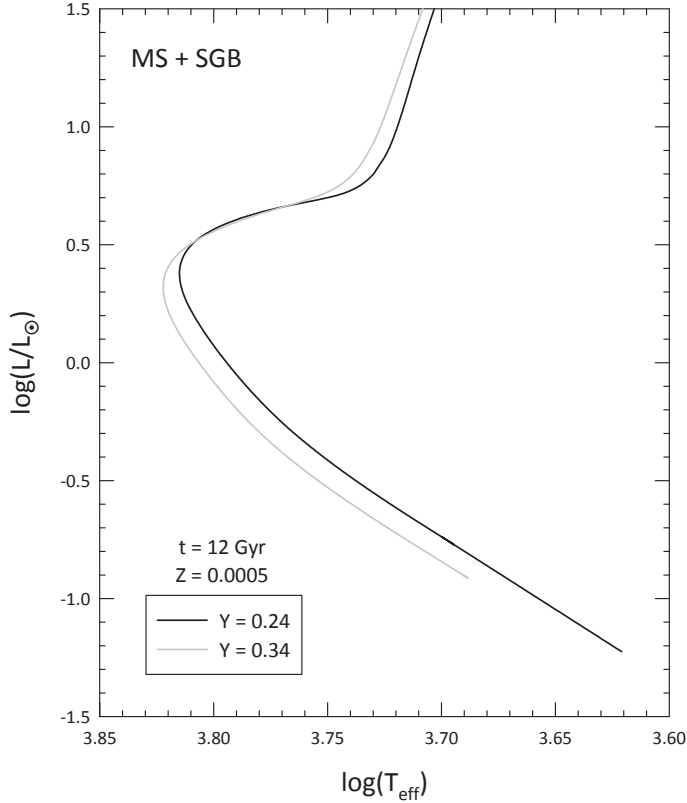} 
 \caption{As in Figure~\ref{fig:HRD}, but zooming in around the MS-SGB region.}
   \label{fig:MS-SGB}
\end{center}
\end{figure}

\begin{figure}[ht]
\begin{center}
 \includegraphics[width=3.57in]{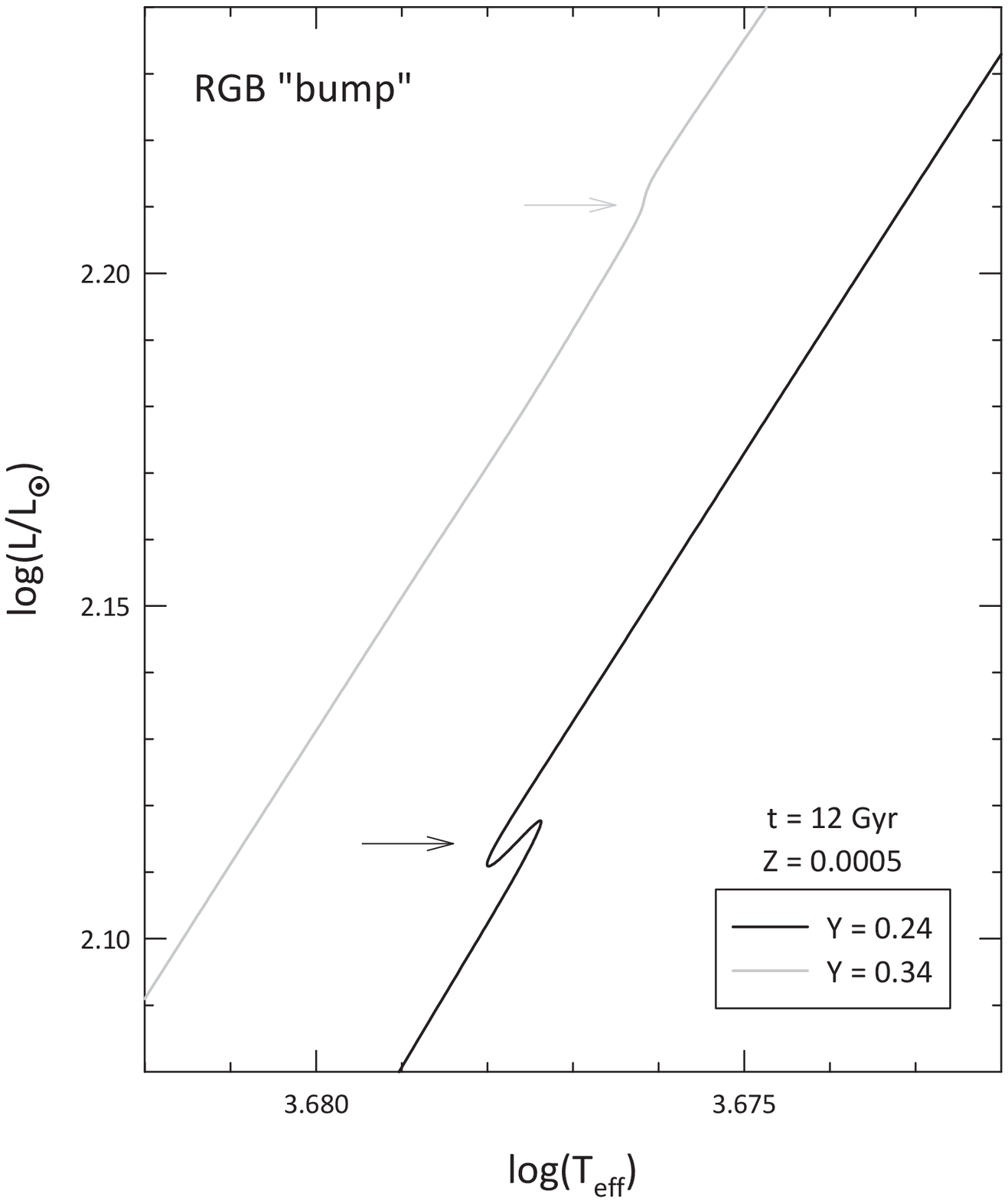} 
 \caption{As in Figure~\ref{fig:HRD}, but zooming in around the RGB bump region.}
   \label{fig:RGB-bump}
\end{center}
\end{figure}

\begin{figure}[ht]
\begin{center}
 \includegraphics[width=3.57in]{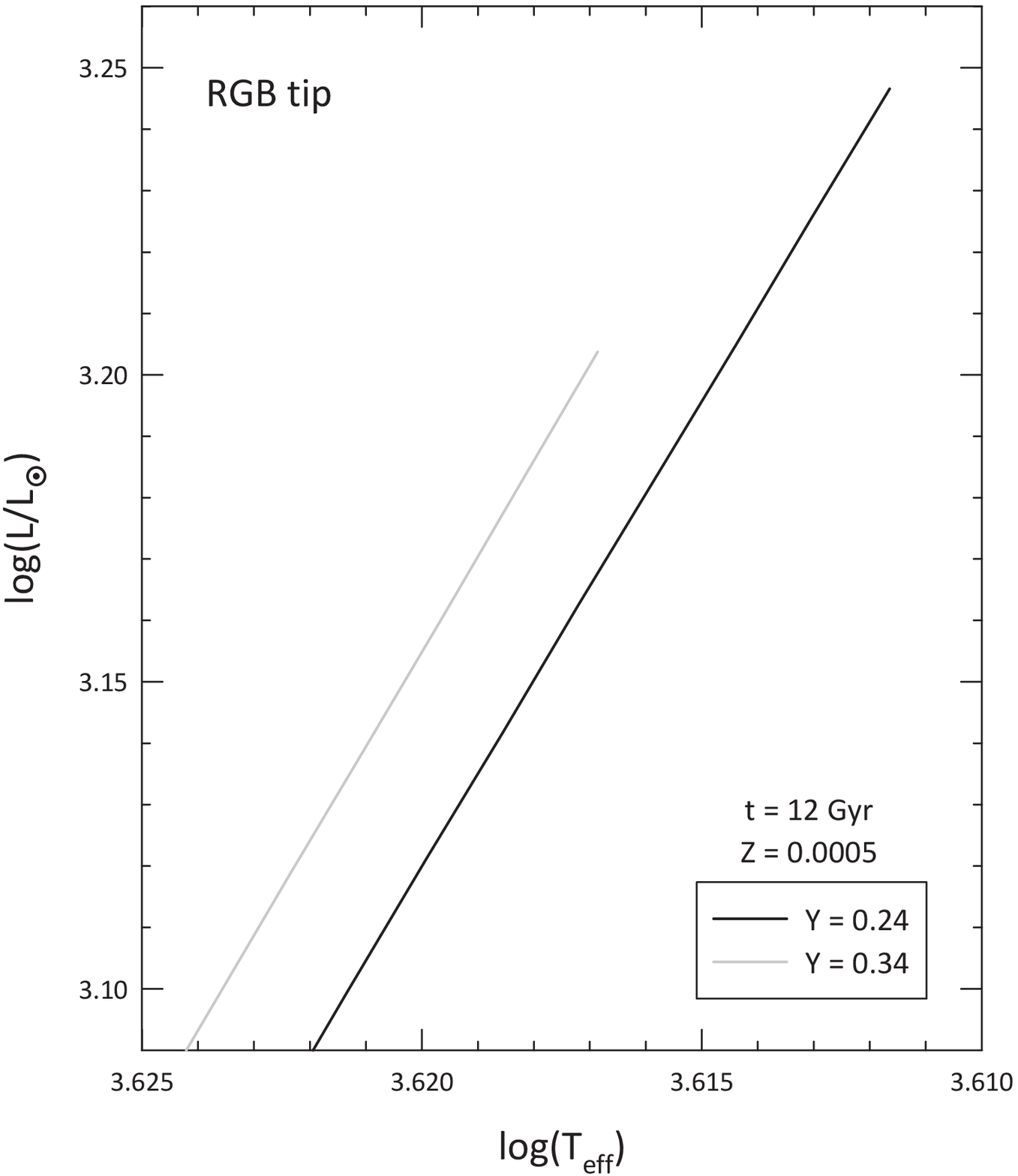} 
 \caption{As in Figure~\ref{fig:HRD}, but zooming in around the RGB tip region.}
   \label{fig:RGB-tip}
\end{center}
\end{figure}

\subsection{Indirect Methods}\label{sec:indirect} 
In Figure~\ref{fig:HRD}, we show a comparison between two isochrones for an old (12~Gyr)
and metal-poor ($Z = 0.0005$) population, for two different He abundances on the zero-age
MS: $Y = 0.24$ (which should be representative of the ``first generation'' of GC stars) 
and $Y = 0.34$ (which is even lower than the amount of He enhancement that has been
suggested for some GC's; see, e.g., Table~1 in \citeauthor{dac08} \citeyear{dac08}). These 
models were computed using an updated version of the Schwarzschild stellar evolution code, 
with updated input physics as described in \citet{av10}. 
The tracks assume a solar-scaled mix (and thus 
${\rm [M/H]} = -1.54$ for these models), but the scaling relation of \citet{msea93} can
be straightforwardly used to apply our tracks to the $\alpha$-enhanced case as well. 

As can be seen from Figure~\ref{fig:HRD}, these isochrones reveal that not only is the MS 
affected by an increase in the He abundance, but so are several other key evolutionary 
stages in the lives of low-mass stars. 
This is summarized in Table~\ref{tab:y024} (for 
a $Y = 0.24$) and Table~\ref{tab:y034} (for a $Y = 0.34$), where the properties of the 
following evolutionary stages are indicated: 
1)~MS, at a temperature $T_{\rm eff} = 5000$~K (corresponding to a color $B\!-\!V \simeq 0.8$); 
2)~MS, at a luminosity $\log (L/L_{\odot}) = -0.5$; 
3)~MS turnoff (TO) point; 
4)~SGB, corresponding to a point in the isochrone that is cooler than the MS TO by $\Delta\log T_{\rm eff} = 0.03$; 
5)~SGB, at a temperature $\log T_{\rm eff} = 3.78$; 
6)~Base of the RGB, corresponding to a point in the isochrone that is brighter than the MS TO 
by $\Delta\log (L/L_{\odot}) = +0.5$;
7)~Base of the RGB, at a luminosity $\log (L/L_{\odot}) = +1.0$; 
8)~RGB luminosity function ``bump''; 
9)~RGB tip; 
10)~Zero-age HB (ZAHB), at a temperature $\log T_{\rm eff} = 3.85$ (which is close to the blue edge of the RR Lyrae instability strip); 
11)~Asymptotic giant branch (AGB) ``clump.''   
The differences between the predicted values for the two different He abundances are given 
in Table~\ref{tab:diffs}, and the implied slopes in Table~\ref{tab:slopes}. 
In what follows, we address some of the outstanding results of this analysis
\citep[see also][]{msea06}.

\begin{table}
\begin{minipage}[b]{0.5\linewidth}\centering
  \begin{center}
  \caption{Evolutionary Predictions for a $Y = 0.24$.$^1$}
  \label{tab:y024}
 {\scriptsize
  \begin{tabular}{|l|c|c|c|}\hline 
{\bf Stage} & {\bf $M/M_{\odot}$} & {\bf $\log T_{\rm eff}$} & {\bf $\log(L/L_{\odot})$} \\\hline 
MS$^2$      & $0.606$            &  $3.699$                 & $-0.744$                  \\
MS$^3$      & $0.661$            &  $3.738$                 & $-0.500$                  \\
TO          & $0.815$            &  $3.815$                 & $ 0.379$                  \\
SGB$^4$     & $0.832$            &  $3.785$                 & $ 0.622$                  \\
SGB$^5$     & $0.833$            &  $3.780$                 & $ 0.636$                  \\
Base RGB$^6$& $0.838$            &  $3.725$                 & $ 0.879$                  \\
Base RGB$^7$& $0.839$            &  $3.719$                 & $ 1.000$                  \\ 
RGB bump    & $0.843$            &  $3.679$                 & $ 2.112$                  \\
RGB tip     & $0.844$            &  $3.611$                 & $ 3.247$                  \\
ZAHB$^8$    & $0.663$            &  $3.850$                 & $ 1.650$                  \\
AGB clump   & $0.663$            &  $3.708$                 & $ 2.012$                  \\\hline
  \end{tabular}
  }
 \end{center}
\vspace{1mm}
 \scriptsize{
 {\it Notes:}\\
  $^1$For an assumed $Z = 0.0005$, $t = 12$~Gyr.\\
  $^2$For a temperature $T_{\rm eff} = 5000$~K.\\ 
  $^3$For a luminosity $\log (L/L_{\odot}) = -0.50$.\\ 
  $^4$At a temperature cooler than the TO by $\Delta\log T_{\rm eff} = 0.03$.\\
  $^5$At a temperature $\log T_{\rm eff} = 3.780$.\\
  $^6$At a level more luminous than the TO by $\Delta\log (L/L_{\odot}) = 0.5$.\\
  $^7$At a luminosity $\log (L/L_{\odot}) = 1.0$.\\
  $^8$At a temperature $\log T_{\rm eff} = 3.850$.}
\end{minipage} 
\begin{minipage}[b]{0.5\linewidth}\centering
  \begin{center}
  \caption{Evolutionary Predictions for a $Y = 0.34$.$^1$}
  \label{tab:y034}
 {\scriptsize
  \begin{tabular}{|l|c|c|c|}\hline 
{\bf Stage} & {\bf $M/M_{\odot}$} & {\bf $\log T_{\rm eff}$} & {\bf $\log(L/L_{\odot})$} \\\hline 
MS$^2$      & $0.513$            &  $3.699$                 & $-0.849$                  \\
MS$^3$      & $0.577$            &  $3.754$                 & $-0.500$                  \\
TO          & $0.687$            &  $3.822$                 & $ 0.321$                  \\
SGB$^4$     & $0.699$            &  $3.792$                 & $ 0.591$                  \\
SGB$^5$     & $0.700$            &  $3.780$                 & $ 0.630$                  \\
Base RGB$^6$& $0.704$            &  $3.737$                 & $ 0.821$                  \\
Base RGB$^7$& $0.705$            &  $3.727$                 & $ 1.000$                  \\
RGB bump    & $0.708$            &  $3.676$                 & $ 2.211$                  \\
RGB tip     & $0.708$            &  $3.617$                 & $ 3.203$                  \\
ZAHB$^8$    & $0.625$            &  $3.850$                 & $ 1.770$                  \\
AGB clump   & $0.625$            &  $3.724$                 & $ 2.157$                  \\\hline
  \end{tabular}
  }
 \end{center}
\vspace{1mm}
 \scriptsize{
 {\it Notes:}\\
  $^1$For an assumed $Z = 0.0005$, $t = 12$~Gyr.\\
  $^2$For a temperature $T_{\rm eff} = 5000$~K.\\ 
  $^3$For a luminosity $\log (L/L_{\odot}) = -0.50$.\\ 
  $^4$At a temperature cooler than the TO by $\Delta\log T_{\rm eff} = 0.03$.\\
  $^5$At a temperature $\log T_{\rm eff} = 3.780$.\\
  $^6$At a level more luminous than the TO by $\Delta\log (L/L_{\odot}) = 0.5$.\\
  $^7$At a luminosity $\log (L/L_{\odot}) = 1.0$.\\
  $^8$At a temperature $\log T_{\rm eff} = 3.850$.}
\end{minipage} 
\end{table}

\begin{table}
\begin{minipage}[b]{0.5\linewidth}\centering
  \begin{center}
  \caption{Differences between Evolutionary Predictions.$^1$}
  \label{tab:diffs}
 {\scriptsize
  \begin{tabular}{|l|c|c|c|}\hline 
{\bf Stage} & {\bf $\Delta(M/M_{\odot})$} & {\bf $\Delta\log T_{\rm eff}$} & {\bf $\Delta\log(L/L_{\odot})$} \\\hline 
MS$^2$      & $-0.093$ & ---      & $-0.105$ \\
MS$^3$      & $-0.084$ & $+0.016$ &  ---     \\
TO          & $-0.128$ & $+0.007$ & $-0.058$ \\
SGB$^4$     & $-0.133$ & $+0.007$ & $-0.033$ \\
SGB$^5$     & $-0.133$ & ---      & $-0.006$ \\
Base RGB$^6$& $-0.134$ & $+0.012$ & $-0.058$ \\
Base RGB$^7$& $-0.134$ & $+0.008$ &  ---     \\
RGB bump    & $-0.135$ & $-0.003$ & $+0.099$  \\
RGB tip     & $-0.136$ & $+0.006$ & $-0.044$ \\
ZAHB$^8$    & $-0.038$ & ---      & $+0.120$  \\
AGB clump   & $-0.038$ & $+0.016$ & $+0.145$  \\\hline
  \end{tabular}
  }
 \end{center}
\vspace{1mm}
 \scriptsize{
 {\it Notes:}\\
  $^1$In the sense $Y = 0.34$ minus $Y = 0.24$, for an assumed $Z = 0.0005$, $t = 12$~Gyr.\\
  $^2$For a temperature $T_{\rm eff} = 5000$~K.\\ 
  $^3$For a luminosity $\log (L/L_{\odot}) = -0.50$.\\ 
  $^4$At a temperature cooler than the TO by $\Delta\log T_{\rm eff} = 0.03$.\\
  $^5$At a temperature $\log T_{\rm eff} = 3.780$.\\
  $^6$At a level more luminous than the TO by $\Delta\log (L/L_{\odot}) = 0.5$.\\
  $^7$At a luminosity $\log (L/L_{\odot}) = 1.0$.\\
  $^8$At a temperature $\log T_{\rm eff} = 3.850$.}
\end{minipage} 
\begin{minipage}[b]{0.5\linewidth}\centering
  \begin{center}
  \caption{Predicted Slopes.$^1$}
  \label{tab:slopes}
 {\scriptsize
  \begin{tabular}{|l|c|c|c|}\hline 
{\bf Stage}  & {\bf $\frac{d\log T_{\rm eff}}{dY}$} & {\bf $\frac{d\log(L/L_{\odot})}{dY}$} & {\bf $\frac{dM_{\rm bol}}{dY}$} \\\hline 
MS$^2$       &  ---                     & $-1.05$   & $+2.63$                  \\
MS$^3$       &  $+0.16$                 & ---       & ---                       \\
TO           &  $+0.07$                 & $-0.58$   & $+1.45$                  \\
SGB$^4$      &  $+0.07$                 & $-0.33$   & $+0.83$                  \\
SGB$^5$      &  ---                     & $-0.06$   & $+0.15$                  \\
Base RGB$^6$ &  $+0.12$                 & $-0.06$   & $+0.15$                  \\
Base RGB$^7$ &  $+0.08$                 & ---       & ---                       \\
RGB bump     &  $-0.03$                 & $+0.99$   & $-2.48$                  \\
RGB tip      &  $+0.06$                 & $-0.44$   & $+1.10$                  \\
ZAHB$^8$     &  ---                     & $+1.20$   & $+3.00$                  \\
AGB clump    &  $+0.16$                 & $+1.45$   & $+3.63$                  \\\hline
  \end{tabular}
  }
 \end{center}
\vspace{1mm}
 \scriptsize{
 {\it Notes:}\\
  $^1$In the sense $Y = 0.34$ minus $Y = 0.24$, for an assumed $Z = 0.0005$, $t = 12$~Gyr.\\
  $^2$For a temperature $T_{\rm eff} = 5000$~K.\\ 
  $^3$For a luminosity $\log (L/L_{\odot}) = -0.50$.\\ 
  $^4$At a temperature cooler than the TO by $\Delta\log T_{\rm eff} = 0.03$.\\
  $^5$At a temperature $\log T_{\rm eff} = 3.780$.\\
  $^6$At a level more luminous than the TO by $\Delta\log (L/L_{\odot}) = 0.5$.\\
  $^7$At a luminosity $\log (L/L_{\odot}) = 1.0$.\\
  $^8$At a temperature $\log T_{\rm eff} = 3.850$.}
\end{minipage} 
\end{table}

\noindent {\bf MS}. An expanded view of Figure~\ref{fig:HRD} around the MS and SGB 
regions is given in Figure~\ref{fig:MS-SGB}, clearly confirming that He-enhanced models 
are hotter and/or less luminous than their low-$Y$ counterparts. Indeed, detections of 
MS splits, as mentioned in \S\ref{sec:intro}, have so far been the main indicator of 
variations in the He abundance within individual GC's. 

\noindent {\bf MS TO}. According to our results, the MS TO of a He-enhanced population
is both hotter and fainter than is the case for a low $Y$, as can also be seen from 
Figure~\ref{fig:MS-SGB}. 

\noindent {\bf SGB}. Our calculations reveal that the position of the SGB is {\em not} 
sensitive to $Y$ (see also Fig.~\ref{fig:MS-SGB}). Differences in He abundance
are thus not a viable candidate to explain the SGB splits that have been found in the 
literature, and changes in the abundances of other abundant species, such as the CNO 
elements, are accordingly also required to reproduce these splits 
\citep[see also][]{scea08,msea08,pvea09}. 

\noindent {\bf Base of the RGB}. Figure~\ref{fig:MS-SGB} shows that the base of the 
RGB of He-enhanced models is hotter (at a given luminosity) than their low-$Y$ 
counterparts. The detected difference corresponds to a 
$d\Delta(B\!-\!V)_{\rm RGB}/dY \approx d\Delta(V\!-\!I)_{\rm RGB}/dY \approx -0.35$, 
implying a $d\Delta(B\!-\!I)_{\rm RGB}/dY \approx -0.7$. 
This is not a negligible effect; as a matter of fact, Table~\ref{tab:diffs} shows
that the temperature split is predicted to be about half as large as the MS separation. 
Our calculations further reveal that such a difference in temperature, hence color, 
persists until the RGB tip, as can also be seen from Figures~\ref{fig:RGB-bump} 
and \ref{fig:RGB-tip}~-- though the size of the predicted split decreases progressively 
towards the RGB tip, where the presence of AGB stars may also complicate any empirical 
tests \citep[see also][]{cd05,msea06,apea09}. 
Still, since photometry of RGB stars is nowadays often precise to a level much better 
than 0.01~mag, large He enhancements might also manifest themselves as RGB splits 
in well-populated CMDs. On the other hand, one should also keep in mind the possibility 
that differences in color transformations between He-enriched and non-He-enriched GC 
stars might mask, at least in part, this effect \citep[but see][]{lgea07}. Indeed, 
such RGB splits appear to have so far been reported only in the 
cases of M4 \citep[NGC~6121;][]{amea08} and NGC~1851 \citep{jwlea09}, and even in those
cases, only using bluer passbands and the Str\"omgren filter system, respectively. 
This suggests that these RGB splits may in reality be due to the effect of abundance 
variations on the color transformations, rather than to a He enhancement proper 
\citep[see also][]{dyea08,dyea09}~-- but further studies would be of much interest. 

\noindent {\bf RGB bump}. Our calculations confirm many previous indications that
both the position and shape of the RGB bump depend sensitively on the He abundance
\citep[e.g.,][]{msea06}. 
This is particularly clear from Figure~\ref{fig:RGB-bump}, which shows that the 
higher-$Y$ models have a brighter bump than do the lower-$Y$ models. Accordingly, 
multiple generations with widely different He abundances should also manifest themselves
in the form of multiple RGB bumps. A search for these multiple bumps has been conducted 
in the case of $\omega$~Cen \citep{asea05,ff09}, with an indication that the implied range 
in $Y$ is smaller than indicated by analyses of the MS splits. However, the 
analysis of this cluster is complicated by the fact that it contains a large spread 
in metallicities and ages, both of which also affect the position of the RGB bump. 
Therefore, further analysis of this and other (monometallic) GC's would certainly 
prove of interest \citep[see also][]{mrea03,cd05,ecea07}. 

\citet{gbea01} carried out a comparison between the number of 
stars around and below the detected bump in a large and homogeneous sample of GC's and
the prediction of He-enriched models. Their results, as summarized in their Figure~2, 
again do not reveal statistically significant evidence for large He abundance variations, 
except perhaps in the case of NGC~6441~-- which however they suggest may be an artifact
of differential reddening, noting that NGC~6441's ``twin,'' NGC~6388, appears entirely 
consistent with the expectations for a ``normal'' $Y$. Interestingly, \citeauthor{gbea01}
also call attention to a possible increase in $Y$ in the blue-HB cluster M13 (NGC~6205), 
but the measurements for this cluster are still consistent with a ``normal'' $Y$, within
the error bars.  

\noindent {\bf RGB tip}. As is clear from Figure~\ref{fig:RGB-tip}, the luminosity of
the RGB tip, where He ignition occurs, also depends on the He abundance, 
by an amount that is slightly less than 50\% of that seen (at a fixed $T_{\rm eff}$) 
on the MS. Therefore, GC's that lack first-generation 
stars and possess only second- and third-generation stars, as has been suggested for GC's
with completely blue HB's \citep[e.g.,][]{dac08}, 
should have fainter RGB tips, by an amount of order 0.11~mag 
for a difference in $Y$ of 0.1 (Table~\ref{tab:diffs}). While evolution close 
to the RGB tip is quite fast, and therefore not many stars are usually found in that 
region of the CMD (which, in addition, may also contain AGB stars), 
careful analysis based on large samples might  
give indications of whether fainter RGB tips do indeed tend to be more common, 
statistically speaking, in blue-HB GC's. In this sense, some authors have used the 
RGB tip luminosity to place constraints on the He core mass at the He flash 
$M_{\rm c}^{\rm HeF}$, which is of strong interest for particle physicists in 
particular. However, such studies have so far revealed evidence for large increases 
in neither $M_{\rm c}^{\rm HeF}$ nor $Y$ with respect to canonical values~-- on the
contrary, available analyses have tended to favor a somewhat low He abundance 
\citep[see, e.g., Fig.~2 in][and references therein]{gr00}, even though more detailed 
analysis of individual clusters, which has not been the focus of most such studies, 
would certainly prove of interest. 

As we have seen, contrary to the RGB tip, the RGB bump  becomes {\em more} luminous
with an increase in $Y$. Therefore, the difference in magnitude between the
RGB tip and the RGB bump should be more sensitive to $Y$ than either of these
features alone. Based on Tables~\ref{tab:y024} and \ref{tab:y034}, we 
find $d\Delta M_{\rm bol}({\rm RGB\,\, bump}-{\rm tip})/dY = -3.6$~-- which is a stronger $Y$
dependence than for many of the individual indicators listed in Table~\ref{tab:slopes}. 

\begin{figure}[ht]
\begin{center}
 \includegraphics[width=3.57in]{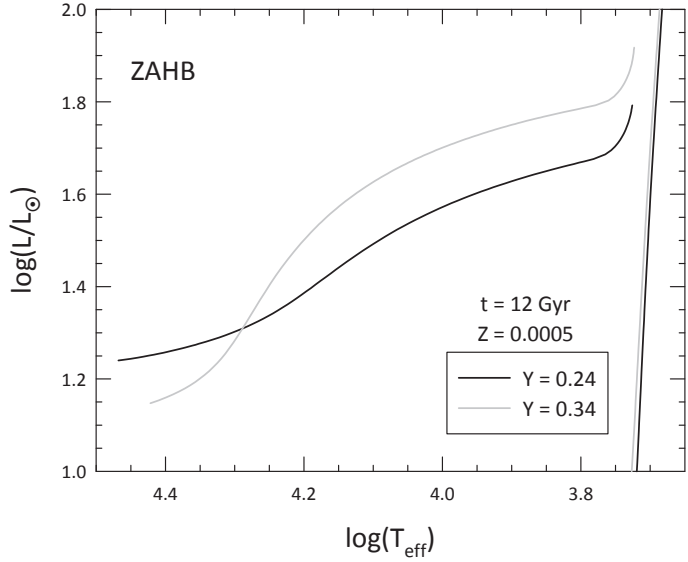} 
 \caption{As in Figure~\ref{fig:HRD}, but zooming in around the ZAHB region.}
   \label{fig:ZAHB}
\end{center}
\end{figure}

\begin{figure}[ht]
\begin{center}
 \includegraphics[width=3.57in]{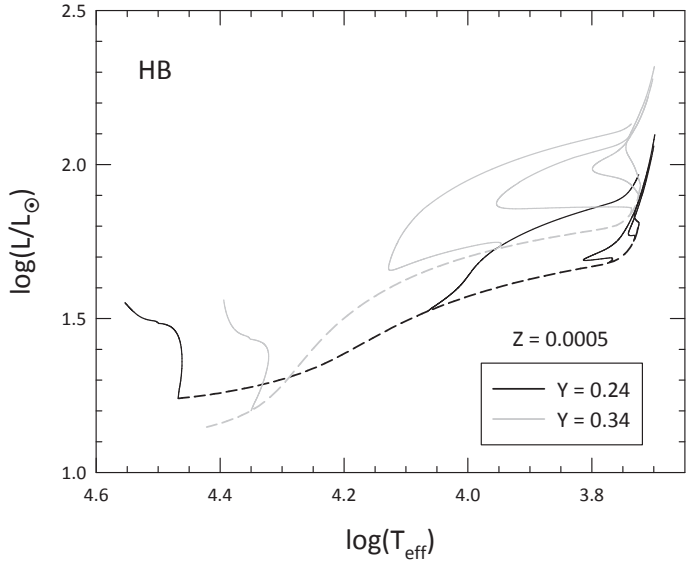} 
 \caption{As in Figure~\ref{fig:ZAHB}, but including evolutionary tracks for the following
   mass values ({\em from left to right}): $0.49\,M_{\odot}$, $0.60\,M_{\odot}$, $0.70\,M_{\odot}$,
   $0.80\,M_{\odot}$, and $0.90\,M_{\odot}$. The same ZAHB sequences as in the previous figure are
   here shown as dashed lines.}
   \label{fig:HB}
\end{center}
\end{figure}

\noindent {\bf ZAHB}. As well known, the ZAHB luminosity is strongly sensitive to $Y$ 
\citep[see, e.g.,][and also our Figs.~\ref{fig:HRD}, \ref{fig:ZAHB}, \ref{fig:HB}]{as87}. 
This strong dependence constitutes the basis for a number 
of methods that are often used to infer the He abundance in GC's. 

One such method is the ``$A$-method'' of \citet{cc75}, which uses the fact that the 
periods of RR Lyrae stars depend strongly on their luminosities, and the latter on $Y$, to 
infer the He abundance on the basis of period and temperature measurements for RR Lyrae
variables. Related to this are the so-called ``period-shift techniques,'' in which 
differences in period, at a given temperature, are taken as evidence for variations in 
luminosity, and hence $Y$. 
Based on these techniques, an overluminosity, and hence evidence
for an increase in $Y$, could not be detected among either $\omega$~Cen's \citep{asea06}
or NGC~1851's \citep{jwlea09} RR Lyrae stars. Similarly, \citet{es00} could not confirm 
the presence of significant He abundance variations in his analysis of a large number 
of GC's, although he did find a difference between metal-poor and moderately metal-rich 
objects, concluding however that ``it is unlikely that the difference in $\langle A\rangle$ 
between the two groups is due to a difference in helium abundance.'' 
On the other hand, large overluminosities, 
possibly related to He enhancement, have been confirmed in the case of the metal-rich 
bulge GC's NGC~6388 and NGC~6441, which are known to contain large populations of RR Lyrae
and blue HB stars~-- an uncommon feature in the metal-rich domain, which by itself may 
also point to the need for He enhancement \citep[e.g.,][]{sc98,bpea02,cd07,gbea07}.

Another method which exploits the strong dependence of HB luminosities on $Y$ is the 
``$\Delta$-method'' of \citet{fcea83}, which uses the difference in magnitude between the 
HB and the MS as a He indicator. This is a particularly strong $Y$ 
indicator because, while the HB luminosity increases with $Y$, the MS luminosity 
decreases; based on Tables~\ref{tab:y024} and \ref{tab:y034}, 
we find a $d\Delta M_{\rm bol}({\rm MS}-{\rm ZAHB})/dY = 5.6$. 
\citet{es00} applied this method
to a large sample of GC's, but did not find evidence for significant variations in $Y$. 

Finally, the increased luminosities of He-enhanced HB stars also lead to decreased 
surface gravities, at a given $T_{\rm eff}$. Therefore, and as suggested by several
authors \citep[see, e.g.,][and references therein]{mc09},
spectroscopic $\log g$ measurements 
can also be of interest in constraining the amount of He enhancement, particularly among 
blue HB stars cooler than 11,500~K, 
whose atmospheres are not affected by the strong diffusion effects that make it extremely 
difficult to properly interpret the results for hotter HB stars. In this sense, the method
was recently applied to the cases of the GC's M3 (NGC~5272) and M13 by \citet{mcea09}, but
no evidence for an increase in $Y$ over the canonical value could be found for either 
cluster. In fact, these authors found, on the basis of a detailed analysis of Str\"omgren 
photometry for M3, that the amount of He enhancement among the cluster's blue HB stars is 
likely less than 0.01 in $Y$, thus ruling out the much higher He enhancements 
that have been proposed in the literature.    

\noindent {\bf HB}. 
The color distribution of HB stars has been widely used, in recent years, to study the 
level of He enhancement in GC's \citep[e.g.,][and references therein]{dac08}. The main 
reason why high-$Y$ GC's of a given age and metallicity are expected to have a bluer HB 
morphology is the fact that their more compact progenitors evolve faster~-- and thus, 
at a given age, high-$Y$ GC's feed stars of lower mass (and thus bluer; see Fig.~\ref{fig:HB})
onto the HB phase than do their lower-$Y$ counterparts. However, the colors of HB stars are
sensitive not only to the He abundance, but also to many other parameters, such as age, 
CNO and $\alpha$-element abundances, rotation, and~-- crucially~-- mass loss on the RGB 
\citep[see][for a recent review]{mc09}. It is thus important to use indicators other 
than the color distribution of HB stars in order to establish whether, and to what level, 
He may be enhanced among HB stars. In this sense, the {\em luminosity distribution} of HB 
stars also plays a very important role that should not be ignored 
\citep[e.g.,][]{ffpea96,dcea88,sc98,msea06,msea08,mcea09}. 

Examination of Figure~\ref{fig:HB} reveals that not only ZAHB luminosities (and their
run with $T_{\rm eff}$), but also other
HB features, are sensitive to the He abundance. Note, in particular, that high-$Y$ 
evolutionary tracks tend to evolve much more significantly in luminosity than their lower-$Y$ 
counterparts, which leads to an increase in the HB ``thickness'' (or luminosity width) 
with increasing $Y$. Similarly, an increase in $Y$ leads to much more pronounced ``blueward
loops,'' which also has important observational consequences 
\citep[e.g.,][]{sc98,bpea02,mc09}. In addition, the predicted 
separation in luminosity between extreme HB (EHB) stars and post-EHB stars is also 
a strong function of $Y$ \citep{tbea08}, which could, at least in principle, provide 
a test of He enhancement for those clusters for which extensive UV photometry is available. 

Another key He abundance indicator for GC's that is based on properties of HB stars is 
the number ratio $R$ between HB stars and RGB stars brighter than the HB \citep[][]{ii68}. 
In short, not only does the HB luminosity increase markedly with increasing $Y$, but also 
HB lifetimes~-- and thus the number of HB stars relative to RGB stars brighter than the 
HB~-- increase markedly with increasing $Y$. Indeed, for HB stars beginning their evolution 
close to the instability strip, our models give a $dR/dY = 8.3$,\footnote{This decreases down
to $dR/dY = 5.8$, if we assume that the HB luminosity taken as reference for RGB number counts
is the same for both He-rich and He-poor models, as might be more appropriate in the case of
clusters for which stars on the ``horizontal'' part of the HB are not He-enhanced~-- for 
which, in addition, one should consider that a fraction of the RGB stars might also be poor
in He \citep[see also][]{mrea03,msea04}.} 
thus confirming the strong sensitivity of the method to He abundance variations. 

The most recent application of this method to a large sample of GC's is the one by 
\citet{msea04}. They find that ``the mean... $Y = 0.250\pm 0.006$. 
An instrinsic dispersion with a firm upper limit of 0.019 around this value... is a priori
possible given the observational errors.'' Interestingly, the authors find some evidence for
higher $R$ values for clusters with blue HB's, as would be expected in the He enhancement 
scenario~-- but point out that this result may be more straightforwardly explained in terms
of the increase in HB lifetimes for bluer, less massive HB stars \citep[see also][]{mzea00}.

\noindent {\bf AGB clump}. According to our results (e.g., Table~\ref{tab:diffs}), the 
AGB clump, which is seen in some GC's at the base of the AGB, 
becomes significantly brighter with increasing $Y$. 
The difference in luminosity between the AGB clump and the ZAHB is, however, 
much less strongly dependent on $Y$, with  
$dM_{\rm bol}({\rm ZAHB}-{\rm AGB\,\, clump})/dY \approx 0.3$ \citep[see also][]{gbea95}. In 
addition, AGB clumps are only found in GC's with predominantly red HB's, which are not 
expected to be the ones displaying the largest levels of He enhancement. 

\firstsection 
\section{Conclusions}\label{sec:conclusions} 
Nature has conspired to make reliable measurements of the He abundance of GC stars 
an extremely difficult task. Still, many features of the CMD, besides MS and HB 
colors, present at least some sensitivity to $Y$~-- which can be best exploited, 
depending on the specific CMD location, in different wavelength regimes, using 
a variety of photometric and spectroscopic techniques. 
It is thus important
to establish, on a cluster-to-cluster basis, which such features are consistent (or 
not), to the limit of our empirical and theoretical knowledge, with the He enhancement 
levels that have been suggested. Only then will we be in a good position to firmly
establish the extent to which such He enhancements occur in Nature. This is 
a crucial task in order to establish the role played by GC's as laboratories of 
stellar evolution~-- and indeed, more generally, of astrophysics.

\firstsection 
\section*{Acknowledgements}
Support for M.C. is provided by Proyecto Basal PFB-06/2007, by FONDAP Centro de Astrof\'{i}sica 15010003, by Proyecto FONDECYT Regular \#1071002, and by a John Simon Guggenheim Memorial Foundation Fellowship. Support for A.V. is provided by IAU, CONICYT, SOCHIAS, MECESUP, and ALMA.

\firstsection 


\begin{thebibliography}{}

\bibitem[Behr(2003)]{bb03}
Behr, B. B. 2003, \apjs, 149, 67

\bibitem[Bono et al.(2001)]{gbea01}
Bono, G., Cassisi, S., Zoccali, M., \& Piotto, G. 2001, \apjl, 546, L109

\bibitem[Bono et al.(1995)]{gbea95}
Bono, G., Castellani, V., Degl'Innocenti, S., \& Pulone, L. 1995, \aap, 297,
115

\bibitem[Brown et al.(2008)]{tbea08}
Brown, T. M., Smith, E., Ferguson, H. C., Sweigart, A. V., Kimble, R. A., \& 
Bowers, C. W. 2008, \apj, 682, 319

\bibitem[Busso et al.(2007)]{gbea07}
Busso, G., et al. 2007, \aap, 474, 105

\bibitem[Caloi \& D'Antona(2005)]{cd05}
  Caloi, V., \& D'Antona, F. 2005, \aap, 435, 987

\bibitem[Caloi \& D'Antona(2007)]{cd07}
  Caloi, V., \& D'Antona, F. 2007, \aap, 463, 949

\bibitem[Caputo \& Castellani(1975)]{cc75}
Caputo, F., \& Castellani, V. 1975, \apss, 38, 39

\bibitem[Caputo et al.(1983)]{fcea83}
Caputo, F., Cayrel, R., \& Cayrel de Strobel, G. 1983, \aap, 123, 135 

\bibitem[Carretta et al.(2009)]{ecea09}
Carretta, E., Bragaglia, A., Gratton, R., \& Lucatello, S. 2009, \aap, in press 
(arXiv:0909.2941)

\bibitem[Carretta et al.(2007)]{ecea07}
Carretta, E., et al. 2007, \aap, 464, 939

\bibitem[Cassisi et al.(2009)]{scea09}
Cassisi, S., Salaris, M., Anderson, J., Piotto, G., Pietrinferni, A., Milone, A., 
Bellini, A., \& Bedin, L. R. 2009, \apj, 702, 1530

\bibitem[Cassisi et al.(2008)]{scea08}
Cassisi, S., Salaris, M., Pietrinferni, A., Piotto, G., Milone, A. P., Bedin, L. R., \& Anderson, J.
2008, \apjl, 672, L115

\bibitem[Catelan(2008)]{mc08}
Catelan, M. 2008, \msait, 79, 388

\bibitem[Catelan(2009)]{mc09}
Catelan, M. 2009, \apss, 320, 261

\bibitem[Catelan et al.(2009)]{mcea09}
Catelan, M., Grundahl, F., Sweigart, A. V., Valcarce, A. A. R., \& Cort\'es, C. 2009, \apj, 695, L97

\bibitem[Crocker et al.(1988)]{dcea88}
Crocker, D. A., Rood, R. T., \& O'Connell, R. W. 1988, \apj, 332, 236

\bibitem[D'Antona et al.(2005)]{fdea05}
D'Antona, F., Bellazzini, M., Caloi, V., Fusi Pecci, F., Galleti, S., \& Rood, R. T. 2005, 
\apj, 631, 868

\bibitem[D'Antona \& Caloi(2008)]{dac08}
  D'Antona, F., \& Caloi, V. 2008, \mnras, 390, 693

\bibitem[Davis et al.(2008)]{ddea08}
Davis, D. S., Richer, H. B., Anderson, J., Brewer, J., Hurley, J., Kalirai, J. S., Rich, 
R. M., \& Stetson, P. B. 2008, \aj, 135, 2155

\bibitem[Dupree et al.(2009)]{adea09}
Dupree, A. K., Smith, G. H., \& Strader, J. 2009, \aj, in press (arXiv:0909.1558)

\bibitem[Ferraro(2009)]{ff09}
Ferraro, F. R. 2009, private communication  

\bibitem[Ferraro et al.(2006)]{ffea06}
Ferraro, F. R., Valenti, E., Straniero, O., \& Origlia, L. 2006, \apj, 642, 225


\bibitem[Fusi Pecci et al.(1996)]{ffpea96}
Fusi Pecci, F., Bellazzini, M., Ferraro, F. R., Buonanno, R., \& Corsi, C. E. 1996, 
in Formation of the Galactic Halo .... Inside and Out, ASP Conf. Ser., Vol. 92, ed. 
H. Morrison \& A. Sarajedini (San Francisco: ASP), 221

\bibitem[Girardi et al.(2007)]{lgea07}
Girardi, L., Castelli, F., Bertelli, G., \& Nasi, E. 2007, \aap, 468, 657

\bibitem[Gratton et al.(2000)]{rgea00}
Gratton, R. G., Carretta, E., Matteucci, F., \& Sneden, C. 2000, \aap, 358, 671

\bibitem[Grundahl et al.(1999)]{fgea99}
  Grundahl, F., Catelan, M., Landsman, W. B., Stetson, P. B., \& Andersen, M. I. 
  1999, \apj, 524, 242

\bibitem[Iben(1968)]{ii68}
  Iben, I., Jr. 1968, \nat, 220, 143

\bibitem[Lee et al.(2009)]{jwlea09}
Lee, J.-W., Lee, J., Kang, Y.-W., Lee, Y.-W., Han, S.-I., Joo, S.-J., Rey, S.-C., \& 
Yong, D. 2009, \apjl, 695, L78

\bibitem[Marcolini et al.(2009)]{amea09}
Marcolini, A., Gibson, B. K., Karakas, A. I., \& S\'anchez-Bl\'azquez, P. 2009, \mnras,
395, 719

\bibitem[Marino et al.(2008)]{amea08}
Marino, A. F., Villanova, S., Piotto, G., Milone, A. P., Momany, Y., Bedin, L. R., \& 
Medling, A. M. 2008, \aap, 490, 625

\bibitem[Moehler(2001)]{sm01}
Moehler, S. 2001, \pasp, 113, 1162

\bibitem[Moehler et al.(2007)]{smea07}
Moehler, S., Dreizler, S., Lanz, T., Bono, G., Sweigart, A. V., Calamida, A., Monelli, 
M., \& Nonino, M. 2007, \aap, 475, L5

\bibitem[Moehler et al.(2003)]{smea03}
Moehler, S., Landsman, W. B., Sweigart, A. V., \& Grundahl, F. 2003, \aap, 405, 135 

\bibitem[Moehler et al.(2004)]{smea04}
Moehler, S., Sweigart, A. V., Landsman, W. B., \& Dreizler, S. 2009, \apss, 291, 231

\bibitem[Norris(2004)]{jn04}
Norris, J. E. 2004, \apjl, 612, L25

\bibitem[Palmieri et al.(2002)]{rpea02}
Palmieri, R., Piotto, G., Saviane, I., Girardi, L., \& Castellani, V. 2002, \aap, 392, 115

\bibitem[Pietrinferni et al.(2009)]{apea09}
Pietrinferni, A., Cassisi, S., Salaris, M., Percival, S., \& Ferguson, J. W. 2009, \apj, 
697, 275

\bibitem[Piotto(2009a)]{gp09a}
Piotto, G. 2009a, {\em Proceedings of the Intl. Astr. Union}, 4, 233

\bibitem[Piotto(2009b)]{gp09b}
Piotto, G. 2009b, these proceedings

\bibitem[Piotto et al.(2005)]{gpea05}
Piotto, G., et al. 2005, \apj, 621, 777

\bibitem[Piotto et al.(2007)]{gpea07}
Piotto, G., et al. 2007, \apjl, 661, L53

\bibitem[Pritzl et al.(2002)]{bpea02}
  Pritzl, B. J., Smith, H. A., Catelan, M., \& Sweigart, A. V. 2002, \aj, 124, 949; 
  erratum: 2003, \aj, 125, 2752

\bibitem[Raffelt(1996)]{gr96}
  Raffelt, G. G. 1996, Stars as Laboratories for Fundamental Physics:
  The Astrophysics of Neutrinos, Axions, and Other Weakly Interacting Particles 
  (Chicago: University of Chicago Press)

\bibitem[Raffelt(2000)]{gr00}
  Raffelt, G. G. 2000, \textit{Phys. Rep.}, 333, 593

\bibitem[Raffelt(2008)]{gr08}
  Raffelt, G. G. 2008, \textit{Lect. Notes Phys.}, 741, 51
  
\bibitem[Riello et al.(2003)]{mrea03}
Riello, M., et al. 2003, \aap, 410, 553  
  
\bibitem[Salaris et al.(2008)Salaris, Cassisi, \& Pietrinferni]{msea08}
Salaris, M., Cassisi, S., \& Pietrinferni, A. 2008, \apjl, 678, L25

\bibitem[Salaris et al.(1993)Salaris, Chieffi, \& Straniero]{msea93}
  Salaris, M., Chieffi, A., \& Straniero, O. 1993, \apj, 414, 580

\bibitem[Salaris et al.(2004)]{msea04}
  Salaris, M., Riello, M., Cassisi, S., \& Piotto, G. 2004, \aap, 420, 911  
  
\bibitem[Salaris et al.(2006)]{msea06}
  Salaris, M., Weiss, A., Ferguson, J. W., \& Fusilier, D. J. 2006, \apj, 645, 1131  
  
\bibitem[Sandquist(2000)]{es00}
Sandquist, E. L. 2000, \mnras, 313, 571  
  
\bibitem[Sollima et al.(2006)]{asea06}
Sollima, A., Borissova, J., Catelan, M., Smith, H. A., Minniti, D., Cacciari, C., \& Ferraro, F. R. 
2006, \apjl, 640, L43

\bibitem[Sollima et al.(2005)]{asea05}
Sollima, A., Ferraro, F. R., Pancino, E., \& Bellazzini, M. 2005, \mnras, 357, 265  
  
\bibitem[Sweigart(1987)]{as87}
  Sweigart, A. V. 1987, \apjs, 65, 95

\bibitem[Sweigart \& Catelan(1998)]{sc98}
  Sweigart, A. V., \& Catelan, M. 1998, \apjl, 501, L63

\bibitem[Valcarce(2010)]{av10}
Valcarce, A. A. R. 2010, Ph.D. Thesis, in preparation

\bibitem[Ventura et al.(2009)]{pvea09}
Ventura, P., Caloi, V., D'Antona, F., Ferguson, J., Milone, A., \& Piotto, G. P. 2009, 
\mnras, in press (arXiv:0907.1765) 

\bibitem[Villanova et al.(2009)]{svea09}
Villanova, S., Piotto, G., \& Gratton, R. 2009, \aap, 499, 755

\bibitem[Yong et al.(2009)]{dyea09}
Yong, D., Grundahl, F., D'Antona, F., Karakas, A. I., Lattanzio, J. C., \& Norris, 
J. E. 2009, \apjl, 695, L62

\bibitem[Yong et al.(2008)]{dyea08}
Yong, D., Grundahl, F., Johnson, J. A., \& Asplund, M. 2008, \apj, 684, 1159

\bibitem[Yoon et al.(2008)]{sjyea08}
Yoon, S.-J., Joo, S.-J., Ree, C. H., Han, S.-I., Kim, D.-G., \& Lee, Y.-W. 2008, \apj, 
677, 1080

\bibitem[Zoccali et al.(2000)]{mzea00}
Zoccali, M., Cassisi, S., Bono, G., Piotto, G., Rich, R. M., \& Djorgovski, 
S. G. 2000, \apj, 538, 289


\end{thebibliography}
\end{document}